\documentclass[twocolumn,pre,preprintnumbers,amsmath,amssymb,floatfix]{revtex4-1}
\usepackage{graphicx}
\usepackage{amsmath}
\usepackage{amssymb}
\usepackage{color}
\begin{document}

\title{Random geometric graph description of connectedness percolation in rod systems}

\author{Avik P. Chatterjee}\affiliation{Department of Chemistry, SUNY College of Environmental Science and Forestry, One Forestry Drive, Syracuse, N.Y. 13210}

\author{Claudio Grimaldi}\affiliation{Laboratory of Physics of Complex Matter, Ecole Polytechnique F\'ed\'erale de
Lausanne, Station 3, CH-1015 Lausanne, Switzerland}

\begin{abstract}
The problem of continuum percolation in dispersions of rods is reformulated in terms of weighted random geometric graphs. 
Nodes (or sites or vertices) in the graph represent spatial locations occupied by the centers of the rods. The probability that 
an edge (or link) connects any randomly selected pair of nodes depends upon the rod volume fraction as well as the distribution 
over their sizes and shapes, and also upon quantities that characterize their state of dispersion (such as the orientational distribution function). 
We employ the observation that contributions from closed loops of connected rods are negligible in the limit of large aspect ratios 
to obtain percolation thresholds that are fully equivalent to those calculated within the second-virial approximation of the 
connectedness Ornstein-Zernike equation. Our formulation can account for effects due to interactions between the rods, and 
many-body features can be partially addressed by suitable choices for the edge probabilities.
\end{abstract}

\maketitle

\section{Introduction}
\label{intro}

Dispersions of rod-like particles in the continuum are prototypical representations of numerous and varied systems 
of diverse and widespread technological interest, including liquid crystals, fiber-reinforced materials, and polymers 
filled with conducting nanofibers and nanotubes. Of particular interest are percolation 
phenomena \cite{Stauffer1994,TorquatoBook} associated with 
long-range connectedness of rod-like particles, as these can dramatically affect the strength of fiber-reinforced 
materials and the electrical conductivity of nanocomposites. 

Traditionally the volume fraction at which an infinite cluster of connected particles first arises (referred to as the 
``percolation threshold") has been studied theoretically by means of integral equation approaches based upon the 
connectedness Ornstein-Zernike equation \cite{Bug1985,Leung1991,Chatterjee2000,Otten2009,Otten2011}. 
For the case of elongated rods that are distributed uniformly and oriented randomly in the matrix, the prediction that 
the percolation threshold depends in an approximately inverse manner upon the aspect ratio, has been confirmed by computer 
simulations for both penetrable and hard-core-soft-shell 
rods \cite{Neda1999,Foygel2005,Berhan2007,Schilling2007,Ambrosetti2010,Mutiso2012,Nigro2013a}. 

In recent years, the impact of polydispersity in the particle size (the fact that in real-life situations, particles 
usually display a distribution over a range of sizes and are not all of uniform dimensions) upon the percolation 
threshold has been examined using both (i) integral equation methods \cite{Otten2009,Otten2011}, and (ii) 
a mapping between continuum percolation and percolation on an appropriately modified, tree-like, Bethe lattice \cite{Chatterjee20102012}. 
The latter formalism applied excluded volume arguments \cite{Balberg1984} to estimate the average number of inter-particle 
contacts as a function of volume fraction, and related these to nearest-neighbor occupancy rates on a Bethe lattice 
with suitably chosen vertex degrees (co-ordination numbers). For cases in which the particles are assumed to have 
polydispersity in their lengths alone and uniform diameters and widths, both of these approaches show that the 
leading-order term determining the percolation threshold varies inversely with the weight-averaged aspect ratio 
of the particles, as confirmed recently by Monte Carlo simulations \cite{Mutiso2012,Nigro2013a,Meyer2015}. The original 
implementations of the lattice analogy neglected fluctuations in the expected number of contacts between particles, 
and led to findings that differed from those obtained from the integral equation methodology for non-leading-order 
terms in the expression for the percolation threshold.

In this work, the continuum percolation problem for rods is reformulated in terms of random geometric graphs with
appropriately specified degree distributions, a vehicle that is 
usually employed to describe connectedness in assemblies of isotropic, spherical objects \cite{Penrose2003,Dall2002,Herrmann2003,Masuda2005,Grimaldi2015}.  
In the present formulation, nodes (or sites or vertices) represent rods that are distributed randomly throughout the matrix, 
and the probability that a randomly chosen pair of nodes is connected by an edge is weighted by the relative orientation of 
the rods. It is shown that the node-degree distribution automatically includes fluctuations in the number of inter-particle 
contacts, and that closed loops of intersecting rods can be neglected in the limit of large aspect ratios. Results for the 
percolation threshold obtained from this formulation are in full agreement with those previously obtained \cite{Bug1985,Otten2009,Otten2011} 
using the methods of integral equation theory. In addition, it is shown that many-body effects for interacting rods and/or non-equilibrium 
features in the distribution of particles can be partially accounted for by suitable choices  for the edge probabilities.

We begin this account by first examining the case of fully penetrable, monodisperse rods in Sec.~\ref{mono}, and use this relatively 
simple situation to develop our notation. Percolation in systems of polydisperse but interpenetrable rods, and percolation in systems of 
rods with interactions (such as pairwise excluded volume), are discussed in Secs.~\ref{poly} and \ref{int}, respectively, and Sec.~\ref{concl} presents 
a summary of and conclusions from this study.

\section{Monodisperse penetrable rods}
\label{mono}

\subsection{Mapping onto a random graph problem: Degree distributions}
\label{mapping}

We start by considering a set of $N$ rod-like particles, the centers of which are randomly located within a three-dimensional 
region of volume $V$. In the interests of achieving a clear presentation, and in order to establish our notation, we start by 
considering first and separately the case of monodisperse rods, which we model as identical penetrable cylinders of length 
$L$ and diameter $\Delta$.  

The orientation of each rod is specified by a unit vector, $\vec{u}$, that is aligned parallel to the axis of the cylinder. 
We assume that each unit vector $\vec{u}$ may take on any one of $M_u\leq N$ possible orientation $\vec{u}_i$ (with $i=1,\,2,\ldots ,M_u$ ) 
that are drawn from a prescribed distribution function. A pair of cylinders with orientations $\vec{u}_i$ and  $\vec{u}_j$ and centered 
at the locations $\vec{r}$ and $\vec{r}\,'$ are said to be connected to each other if the (penetrable) boundaries of the particles 
overlap each other. We next construct a network with $N$ points (or vertices or nodes) by associating a node with the center of 
each particle. Edges (or links) connecting pairs of nodes are assigned based upon whether or not a given pair of particles (or nodes) 
are connected to each other based upon the foregoing criterion for inter-particle overlap. This  $N$-point network may be regarded 
as a weighted random geometric graph, in which the probability of edge formation between any pair of randomly selected points (nodes) 
depends upon the relative positions and orientations of the cylinders that are associated with the selected nodes. We next introduce 
the connectedness function $f_{ij}=f(\vec{R},\vec{u}_i,\vec{u}_j)$, where $\vec{R}=\vec{r}-\vec{r}\,'$ is the displacement vector
between the centers of the pair of cylinders, defined such that $f_{ij}$ is equal to unity if the pair 
of cylinders overlap and is otherwise equal to zero. The probability that the nodes representing particles of orientations $\vec{u}_i$ 
and $\vec{u}_j$ are directly linked by an edge is then given by the integral of $f_{ij}$ over $\vec{R}$:
\begin{equation}
\label{vij1}
\upsilon_{ij}=\frac{1}{V}\int d\vec{R}f(\vec{R},\vec{u}_i,\vec{u}_j).
\end{equation}
The displacement vector can be expressed as 
$\vec{R}=z\vec{u}_i+z'\vec{u}_j+\lambda(\vec{u}_i\times\vec{u}_j)/\vert\vec{u}_i\times\vec{u}_j\vert$ \cite{Straley1973},
where $\lambda$ is the component of  $\vec{R}$ along the direction of the shortest line connecting $i$ and $j$, 
and  $z$ and $z'$ are components along the directions of $\vec{u}_i$  and  $\vec{u}_j$. 
In the limit $L\gg\Delta$, the connectedness function is expressed in terms of $z$, $z'$, and $\lambda$ as simply  
$f_{ij}=\theta(L/2-\vert z\vert)\theta(L/2-\vert z'\vert)\theta(\Delta-\vert\lambda\vert)$ \cite{Otten2011,Straley1973},
where $\theta$ is the Heaviside step function.  
Using $d\vec{R}=\vert\sin\gamma_{ij}\vert dzdz'd\lambda$, where $\gamma_{ij}=\gamma(\vec{u}_i,\vec{u}_j)$ 
is the angle between the directions of $\vec{u}_i$ and $\vec{u}_j$, Eq.~\eqref{vij1} reduces to $\upsilon_{ij}=V_{ij}^\textrm{exc}/V$, where
\begin{equation}
\label{Vex1}
V_{ij}^\textrm{exc}=2L^2\Delta\vert\sin\gamma_{ij}\vert
\end{equation}
is the leading-order contribution to the excluded volume \cite{Onsager1949}. The dependence of  $V_{ij}^\textrm{exc}$ upon the relative orientation of 
the pair of cylinders can be thought of as providing a weight factor for the edge probability that is maximum when the cylinders are orthogonal and 
vanishes for parallel configurations. Note that the vanishing of $\upsilon_{ij}$  when $\gamma_{ij}=0$ and $\pi$  is an artifact of the  $L\gg\Delta$ limit, 
which has negligible impact on the connectivity of slender rods with isotropic orientational distributions.

In order to proceed further, we must introduce the degree distribution function that characterizes the connectedness properties of the network. 
We construct the probability $p_i(k_1,\vec{u}_1;k_2,\vec{u}_2;\ldots)$ that a randomly selected cylinder $i$ with orientation $\vec{u}_i$ is 
connected to $k_1$ cylinders with orientation $\vec{u}_1$, $k_2$ cylinders with orientation $\vec{u}_2$, and so on. Since we have assumed 
that the orientations of different cylinders are independently distributed,  $p_i(k_1,\vec{u}_1;k_2,\vec{u}_2;\ldots)$ is a product of binomial 
distributions $p_{ij}(k_j)$, each of which gives the probability that the cylinder of type $i$ is connected to precisely $k_j$ cylinders of type $j$:
\begin{equation}
\label{distr1}
p_i(k_1,\vec{u}_1;k_2,\vec{u}_2;\ldots)=\prod_j p_{ij}(k_j),
\end{equation}
where
\begin{equation}
\label{distr2}
p_{ij}(k_j)=\binom{N_j-\delta_{ij}}{k_j}\upsilon_{ij}^{k_j}(1-\upsilon_{ij})^{N_j-\delta_{ij}-k_j},
\end{equation}
where $N_j$ is the number of cylinders with orientation vector $\vec{u}_j$, and $\delta_{ij}$ is the Kronecker symbol. 
We next take the limit  $V\rightarrow\infty$ in such a manner that the 
number density $\rho_i=N_i/V$ remains finite for all $i$. Introducing the number fraction $x_i=N_i/N$ and the total number density $\rho=N/V$, 
Eq.~\eqref{distr2} reduces to a Poisson distribution in this limit:
\begin{equation}
\label{distr3}
p_{ij}(k_j)=\frac{z_{ij}^{k_j}}{k_j!}e^{-z_{ij}},
\end{equation}
where
\begin{equation}
\label{z1}
z_{ij}=x_j\rho V_{ij}^\textrm{exc}=2x_j\rho L^2\Delta\vert\sin\gamma_{ij}\vert,
\end{equation}
is the average number of cylinders of type $j$ that overlap a given cylinder of type $i$. The distribution function in Eq.~\eqref{distr3} accounts 
for fluctuations in the number of contacts that may exist between pairs of cylinders of types $i$ and $j$, which is an issue that was neglected 
in a prior effort to model percolation by rods within a lattice-based approach \cite{Chatterjee20102012}. 

Knowledge of the degree distribution of Eq.~\eqref{distr3} is sufficient to fully ascertain the connectivity properties of the system, provided that 
we consider the limit of slender rods (that is, $L\gg\Delta$) and assume that the orientations of the rods are distributed isotropically. This is so 
because under these conditions the probability of finding closed loops in finite clusters of isotropically oriented connected cylinders scales as 
$\Delta/L$  (as shown in the Appendix) and thus vanishes in the limit $L/\Delta\rightarrow\infty$. Consequently, the random geometric graph 
constructed as described above has a tree-like, dendritic, structure, where the number of branches attached to a randomly selected node 
follows the distribution function given by Eqs.~\eqref{distr1} and \eqref{distr3}.

\subsection{Calculation of the percolation threshold }
\label{perc1}

In this section, we calculate the percolation threshold by following a method that takes advantage of the tree-like structure of the graph 
developed in Sec.~\ref{mapping}. We consider a randomly selected edge that connects a node of type $i$ with a node of type $j$, and we
follow that edge from $i$ to $j$. The node of type $j$ that we arrive at by following that edge will be $k_i$ times more likely to 
have degree $k_i$ than degree one in terms of its direct links to nodes of type $i$. The degree distribution for the node
of type $j$ that is arrived at in this way, denoted $q_{ji}(k_1,\vec{u}_1;k_2,\vec{u}_2;\ldots)$, must therefore be proportional
to $k_ip_j(k_1,\vec{u}_1;k_2,\vec{u}_2;\ldots)$ \cite{Newman2001,Albert2002,Leicht2009}. Using Eqs.~\eqref{distr1} and \eqref{distr3} and
requiring that $q_{ji}(k_1,\vec{u}_1;k_2,\vec{u}_2;\ldots)$ be correctly normalized leads to
\begin{align}
\label{distr4}
q_{ji}(k_1,\vec{u}_1;k_2,\vec{u}_2;\ldots)&=\frac{k_i p_{ji}(k_i)}{z_{ji}}\prod_{l\neq i}p_{jl}(k_l)\nonumber \\
&=p_{ji}(k_i-1)\prod_{l\neq i}p_{jl}(k_l).
\end{align} 
Given the dendritic structure of the network, we can write the probability $Q_{ji}$ that none of the remaining other edges leading 
out of the node of type $j$ leads to the the giant component as follows: 
\begin{equation}
\label{Q1}
Q_{ji}=\sum_{k_1,k_2,\ldots}q_{ji}(k_1,\vec{u}_1;k_2,\vec{u}_2;\ldots)\prod_m Q_{mj}^{k_m-\delta_{mi}},
\end{equation}
where the Kronecker symbol $\delta_{mi}$ accounts for the fact that there are $k_i-1$  edges with nodes of type $i$ other than the initially 
selected one that was followed to arrive at the node of type $j$. Equations \eqref{distr4} and \eqref{Q1} lead to
\begin{align}
\label{Q2}
Q_{ji}=&e^{z_{ji}(Q_{ij}-1)}\prod_{l\neq i}\left[\sum_k p_{jl}(k)Q_{lj}^k\right]\nonumber \\
=&\exp\!\left[\sum_l z_{jl}(Q_{lj}-1)\right],
\end{align}
from which we see that $Q_{ji}$ depends only on the first index: $Q_{ji}=Q_j$. It should be noted that Eq.~\eqref{Q2} always admits the trivial 
solution $Q_j=1$, which corresponds to the absence of infinite, connected clusters. The existence of non-trivial solutions such that $Q_j<1$ 
indicates the presence of an infinite cluster (giant component) comprising a finite fraction of nodes, and that spans the entire 
system \cite{Stauffer1994}. The percolation threshold is identified by evaluating the smallest fraction of nodes for which such a 
non-trivial solution of Eq.~\eqref{Q2} exists.  We thus set $Q_j=1-\varepsilon_j$, with $\varepsilon_j\ll 1$, and expand Eq.~\eqref{Q2} to first 
order in $\varepsilon_j$, which yields
\begin{equation}
\label{eps1}
\varepsilon_j=\sum_l z_{jl}\varepsilon_l=2\rho L^2\Delta\sum_l x_l\vert\sin\gamma_{jl}\vert\varepsilon_l ,
\end{equation}
where we have used Eq.~\eqref{z1}. Passing to a continuum representation of the orientational distribution by introducing 
$\rho(\vec{u})=\sum_i x_i \delta(\vec{u}-\vec{u}_i)$ , Eq.~\eqref{eps1} can be written as
\begin{equation}
\label{eps2}
\varepsilon(\vec{u})=2\rho L^2\Delta\int d\vec{u}'\rho(\vec{u}')\vert\sin[\gamma(\vec{u},\vec{u}')]\vert\varepsilon(\vec{u}').
\end{equation}
For the case of an isotropic orientational distribution of the rods, for which $\rho(\vec{u})=1/4\pi$, performing the integral over $\vec{u}$ on both 
sides of Eq.~\eqref{eps2}, we find the following result for the percolation threshold: 
\begin{equation}
\label{pt1}
\pi\rho_c L^2\Delta/2=1,
\end{equation}
where we have used $(4\pi)^{-1}\int d\vec{u}\vert\sin[\gamma(\vec{u},\vec{u}')]\vert=\pi/4$ and $\rho_c$ denotes the number
density at the percolation threshold. Introducing the dimensionless critical density $\eta_c=\rho_c\pi \Delta^2L/4$, 
Eq.~\eqref{pt1} can be rewritten as
\begin{equation}
\label{pt2}
\eta_c=\frac{1}{2}\frac{\Delta}{L},
\end{equation}
which coincides with the percolation threshold of slender rods obtained from connectedness percolation theory within the 
second-virial approximation \cite{Bug1985}.

We have obtained Eq.~\eqref{pt2} by finding the density of rods such that Eq.~\eqref{Q2} has a non-trivial solution (other than the trivial 
solution identically equal to unity). Owing to the absence of correlations between penetrable rods, we could, instead, have followed a 
different procedure by initially choosing a density $\rho'$ for the rods that is (by construction) large enough that an infinite connected cluster 
is certain to exist, and subsequently randomly deleting a given fraction $q$ of rods until the damaged network just barely ceases to remain 
infinitely connected (the Ògiant componentÓ disappears) \cite{Cohen2000}. This alternative procedure is equivalent to rewriting Eq.~\eqref{Q1} as
\begin{equation}
\label{Q3}
Q_{ji}=q+(1-q)\sum_{k_1,k_2,\ldots}q_{ji}(k_1,\vec{u}_1;k_2,\vec{u}_2;\ldots)\prod_m Q_{mj}^{k_m-\delta_{mi}},
\end{equation}
where $q_{ji}(k_1,\vec{u}_1;k_2,\vec{u}_2;\ldots)$ is the distribution of Eq.~\eqref{distr4} for rod density fixed at $\rho'$. By repeating the same 
steps of the above analysis, we find that the critical fraction $q_c$  of deleted nodes satisfies 
\begin{equation}
\label{pt3}
(1-q_c)\pi\rho' L^2\Delta/2=1,
\end{equation}
so that by defining $\eta_c=(1-q_c)\rho'\pi \Delta^2L/4$  we obtain a result identical to that in Eq.~\eqref{pt2}. We note that the above procedure 
of randomly deleting nodes from a preexisting giant component is equivalent to the one followed in Refs.~\cite{Chatterjee20102012} to establish a 
mapping between continuum percolation of rods and percolation on a modified Bethe lattice. In the present approach, however, fluctuations 
in the number of contacts are fully taken into account, as is apparent from Eq.~\eqref{distr3}.

\section{Polydisperse penetrable rods}
\label{poly}

Having established our notation and formalism for the case of monodisperse rods, we next examine situations in which the cylinders 
have a distribution of lengths and diameters. We start by adopting a discrete representation, such that the length and diameter of each 
cylinder may assume the values $L_{i_L}$ and $\Delta_{i_\Delta}$, respectively, with $i_L=1,\,2,\ldots$ and $i_\Delta=1,\,2,\ldots$, with 
probabilities denoted $x_{i_L,i_\Delta}$. The three-component index vector $\mathbf{i}=(i_L,i_\Delta,i_u)$ is introduced to identify the 
length, diameter, and orientation of each cylinder. With this notation, a straightforward generalization of Eq.~\eqref{vij1} shows that the 
probability that a pair of such randomly located and penetrable cylinders are connected is given by 
$\upsilon_{\mathbf{i}\mathbf{j}}=V_{\mathbf{i}\mathbf{j}}^\textrm{exc}/V$, where
\begin{equation}
\label{Vex2}
V_{\mathbf{i}\mathbf{j}}^\textrm{exc}=2L_{i_L}L_{j_L}\left(\frac{\Delta_{i_\Delta}+\Delta_{j_\Delta}}{2}\right)\vert\sin\gamma_{i_u,j_u}\vert
\end{equation}
is the excluded volume for a pair of cylinders with lengths, diameters, and orientations given by $L_{i_L},\Delta_{i_\Delta},\vec{u}_{i_u}$ 
and $L_{j_L},\Delta_{j_\Delta},\vec{u}_{j_u}$, respectively. (The derivation of Eq.~\eqref{Vex2} implicitly assumes that the lengths of the 
cylinders are much larger than their diameters.) An analysis very similar to that presented in Sec.~\ref{mapping} shows that the probability
$p_\mathbf{i}(k_\mathbf{j},\mathbf{j};k_\mathbf{l},\mathbf{l};\ldots)$ that a node of type $\mathbf{i}$ is connected to precisely $k_\mathbf{j}$ 
nodes of type $\mathbf{j}$,  $k_\mathbf{l}$ nodes of type $\mathbf{l}$, and so forth, is
\begin{equation}
\label{distrp1}
p_\mathbf{i}(k_\mathbf{j},\mathbf{j};k_\mathbf{l},\mathbf{l};\ldots)=\prod_\mathbf{j}p_{\mathbf{i}\mathbf{j}}(k_\mathbf{j}),
\end{equation}
with
\begin{equation}
\label{distrp2}
p_{\mathbf{i}\mathbf{j}}(k_\mathbf{j})=\frac{z_{\mathbf{i}\mathbf{j}}^{k_\mathbf{j}}}{k_\mathbf{j}!}e^{-z_{\mathbf{i}\mathbf{j}}},
\end{equation}
where
\begin{align}
\label{zp1}
z_{\mathbf{i}\mathbf{j}}&=x_{j_L,j_\Delta}x_{j_u}\rho V_{\mathbf{i}\mathbf{j}}^\textrm{exc}\nonumber \\
&=2x_{j_L,j_\Delta}x_{j_u}\rho L_{i_L}L_{j_L}
\!\left(\frac{\Delta_{i_\Delta}+\Delta_{j_\Delta}}{2}\right)\vert\sin\gamma_{i_u,j_u}\vert
\end{align}
is the average number of edges that an $\mathbf{i}$ node forms with nodes of type $\mathbf{j}$. Within the \textit{ansatz} of tree-like 
connectedness for the rod network, analysis that follows the steps of Sec.~\ref{perc1} leads to the following analog to Eq.~\eqref{Q2} 
that now includes the index vectors:
\begin{equation}
\label{Q4}
Q_{\mathbf{j}\mathbf{i}}=\exp\left[\sum_\mathbf{l}z_{\mathbf{j}\mathbf{l}}(Q_{\mathbf{l}\mathbf{j}}-1)\right].
\end{equation}
Setting $Q_{\mathbf{j}\mathbf{i}}=Q_\mathbf{j}=1-\varepsilon_\mathbf{j}$, we find the following for $\varepsilon_\mathbf{j}\ll 1$:
\begin{align}
\label{eps3}
\varepsilon_\mathbf{j}&=\sum_\mathbf{l}z_{\mathbf{j}\mathbf{l}}\varepsilon_\mathbf{l}\nonumber \\
&=2\rho\sum_\mathbf{l}x_{l_L,l_\Delta}x_{l_u}L_{j_L}L_{l_L}
\!\left(\frac{\Delta_{j_\Delta}+\Delta_{l_\Delta}}{2}\right)\vert\sin\gamma_{j_u,l_u}\vert\varepsilon_\mathbf{l}.
\end{align}
For an isotropic orientational distribution for the rods, averaging over the angles leads to
\begin{align}
\label{eps4}
\varepsilon_{j_L,j_\Delta}=&\frac{\pi}{4}\rho L_{j_L}\Delta_{j_\Delta}\sum_{l_L,l_\Delta}x_{l_L,l_\Delta}L_{l_L}\varepsilon_{l_L,l_\Delta}\nonumber \\
&+\frac{\pi}{4}\rho L_{j_L}\sum_{l_L,l_\Delta}x_{l_L,l_\Delta}L_{l_L}\Delta_{l_\Delta}\varepsilon_{l_L,l_\Delta},
\end{align}
where $\varepsilon_{j_L,j_\Delta}=\sum_{j_u}x_{j_u}\varepsilon_\mathbf{j}$. Introducing auxiliary variables defined 
as:  $\varepsilon_1=\sum_{j_L,j_\Delta}x_{j_L,j_\Delta}L_{j_L}\varepsilon_{j_L,j_\Delta}$ 
and  $\varepsilon_2=\sum_{j_L,j_\Delta}x_{j_L,j_\Delta}L_{j_L}\Delta_{j_\Delta}\varepsilon_{j_L,j_\Delta}$, Eq.~\eqref{eps4} can be rewritten as
\begin{align}
\label{eps5}
\varepsilon_1&=\frac{\pi}{4}\rho\langle L^2\Delta\rangle\varepsilon_1+\frac{\pi}{4}\rho\langle L^2\rangle\varepsilon_2,\nonumber\\
\varepsilon_2&=\frac{\pi}{4}\rho\langle L^2\Delta^2\rangle\varepsilon_1+\frac{\pi}{4}\rho\langle L^2\Delta\rangle\varepsilon_2,
\end{align}
where $\langle L^2\Delta^m\rangle=\sum_{j_L,j_\Delta}x_{j_L,j_\Delta}(L_{j_L})^2(\Delta_{j_\Delta})^m$ for $m=0,\, 1,\, 2$.
The condition that the linear and homogeneous system of Eq.~\eqref{eps5} has a non-trivial solution may be expressed in terms of the vanishing 
of the appropriate determinant formed from the coefficients, that is, 
\begin{equation}
\label{det}
\left(\frac{\pi}{4}\rho_c\langle L^2\Delta\rangle-1\right)^2-\left(\frac{\pi}{4}\rho_c\right)^2\langle L^2\rangle\langle L^2\Delta^2\rangle=0,
\end{equation}
from which we find that the critical threshold $\eta_c=\pi\rho_c\langle L\Delta^2\rangle/4$ is given by
\begin{equation}
\label{pt4}
\eta_c=\frac{\langle L\Delta^2\rangle}{\langle L^2\Delta\rangle+\sqrt{\langle L^2\rangle\langle L^2\Delta^2\rangle}},
\end{equation}
which coincides with the result obtained from continuum percolation theory based upon the connectedness Ornstein-Zernike equation \cite{Otten2011}. 
Note that for cases in which the rods are assumed to have polydispersity in their lengths alone and identical diameters, Eq.~\eqref{pt4} reduces 
to $\eta_c=\Delta/2L_w$, where $L_w=\langle L^2\rangle/\langle L\rangle$ is the weight average of the rod lengths. It is straightforward to show that 
Eq.~\eqref{pt4} can also be obtained by randomly deleting a fraction $q_c$ of nodes from an infinitely connected cluster until the giant component 
ceases to exist, as was done in Sec.~\ref{perc1} for the case of monodisperse rods.

\section{Interacting rods}
\label{int}

In actual real-life systems containing rod-like particles, such as nanofibers or nanotubes dispersed in polymeric matrices, hard-core potentials 
that are reflected in steric and excluded-volume effects prohibit the intersection of pairs of particles. This consideration limits the practical 
utility of the model of ideal, fully penetrable rods. Furthermore, surface functionalization or addition of depletants can induce effective attractive 
or repulsive forces between the rods that strongly influence the onset of percolation \cite{Schilling2007,Mutiso2015,Nigro2013b,Vigolo2005}.  
In order to illustrate the effects of rod interactions upon connectedness properties within our present formalism of random geometric graphs, 
we consider for simplicity impenetrable monodisperse cylinders with the hard-core diameter $D$, surrounded by a cylindrical penetrable shells of 
thickness $(\Delta-D)/2\geq 0$. As before, we work within the limit where the rods can be treated as being slender or of high aspect ratio, namely, 
that $L\gg\Delta,\,D$. For homogeneous dispersions of rods the probability that a pair of cylinders of orientations $\vec{u}_i$ and
$\vec{u}_j$ are directly linked by an edge is given by
\begin{equation}
\label{vij2}
\upsilon_{ij}=\frac{1}{V}\int d\vec{R}f(\vec{R},\vec{u}_i,\vec{u}_j)g_2(\vec{R},\vec{u}_i,\vec{u}_j),
\end{equation}
where $g_2(\vec{R},\vec{u}_i,\vec{u}_j)$ is the pair distribution function defined such that $V^{-1}g_2(\vec{R},\vec{u}_i,\vec{u}_j)d\vec{R}$
is the probability of finding a rod of orientation $\vec{u}_j$ within the volume element $d\vec{R}$ centered about the position of a 
randomly selected rod of orientation $\vec{u}_i$ \cite{Hansen2006}. For ideal, penetrable cylinders (that is, for $D=0$), $g_2(\vec{R},\vec{u}_i,\vec{u}_j)=1$, 
from which we recover Eq.~\eqref{vij1}. For impenetrable cylinders, $g_2(\vec{R},\vec{u}_i,\vec{u}_j)=0$ whenever the hard cores of the two 
cylinders overlap. In the slender rod limit and for small rod densities, we take the pair distribution function to have the following approximate form:
\begin{equation}
\label{rdf1}
g_2(\vec{R},\vec{u}_i,\vec{u}_j)\simeq\theta(\vert\lambda\vert-D),
\end{equation}
where $\lambda$ is the component of $\vec{R}$ along the direction of the shortest line connecting $i$ and $j$.
The form of $g_2(\vec{R},\vec{u}_i,\vec{u}_j)$ in Eq.~\eqref{rdf1} forbids interpenetration of the cores of the cylinders and assumes that 
for $\vert\lambda\vert>D$ the rods are completely uncorrelated. Using the coordinate system introduced in Sec.~\ref{mapping}, Eqs.~\eqref{vij2} and 
\eqref{rdf1} yield
\begin{equation}
\label{vij3}
\upsilon_{ij}=\frac{2 L^2(\Delta-D)\vert\sin\gamma_{ij}\vert}{V},
\end{equation}
which corresponds to the $L\gg\Delta,\,D$ limit of the excluded volume in units of $V$ for a pair of hard rods with penetrable shells. Following the 
same steps described in Sec.~\ref{mono}, and assuming that the node degree distribution can still be expressed as a product of Poisson distributions 
as in Eqs.~\eqref{distr1} and \eqref{distr2}, we find that the critical volume fraction for the cores of the rods $\phi_c=\pi\rho_c LD^2/4$ at which the 
system first percolates is
\begin{equation}
\label{pt5}
\phi_c=\frac{1}{2}\frac{D^2}{L(\Delta-D)},
\end{equation}
which coincides with the percolation threshold calculated by solving the connectedness Ornstein-Zernike equations for hard-core--soft-shell rods 
in the second-virial approximation \cite{Otten2009,Otten2011}. Analysis performed along lines that are similar to those followed 
in this section and in Sec.~\ref{poly} reveals that, for isotropically oriented polydisperse rods with impenetrable hard cores, our formalism yields 
a result that is identical to that obtained from the connectedness Ornstein-Zernike equations in Ref. \cite{Otten2009}. 

Our description of the impact upon the percolation threshold of addressing inter-particle interactions can be generalized by rewriting 
Eq.~\eqref{rdf1} as follows: 
\begin{equation}
\label{rdf2}
g_2(\vec{R},\vec{u}_i,\vec{u}_j)=\chi(\vert\lambda\vert,\phi)\theta(\vert\lambda\vert-D),
\end{equation}
where $\chi(\vert\lambda\vert,\phi)$ allows for correlations between hard rods even for $\vert\lambda\vert>D$ (for illustrative purposes, the following 
analysis neglects possible dependencies of $\chi$ on the orientation vectors $\vec{u}_i$ and $\vec{u}_j$). In principle, $\chi(\vert\lambda\vert,\phi)$ 
may also describe cases in which impenetrable rods interact through an additional short-range attractive or repulsive potential and/or nonequilibrium 
features of the particle distribution. It is straightforward to show that Eqs.~\eqref{vij2} and \eqref{rdf2} lead to the following implicit equation for the 
critical volume fraction: 
\begin{equation}
\label{pt6}
\chi_\Delta(\phi_c)\phi_c=\frac{1}{2}\frac{D^2}{L(\Delta-D)},
\end{equation}
where $\chi_\Delta(\phi)=(\Delta-D)^{-1}\int_D^\Delta d\lambda\chi(\lambda,\phi)$.
Although knowledge of $\phi_c$ depends on the specific functional form assumed for $\chi_\Delta(\phi)$, Eq.~\eqref{pt6} may nevertheless be 
used to predict the main effects of correlations between rods on the percolation threshold. For example, for hard rods with sufficiently small values 
of $\Delta-D$, we can approximate $\chi_\Delta(\phi)$ by the contact value of the pair distribution function. Using either the Carnahan-Starling 
approximation for hard spheres at the same volume fraction $\chi_\Delta(\phi)=(1-\phi/2)/(1-\phi)^3$, as proposed in 
Refs.~\cite{Franco-Melgar2008,Chatterjee2014}, or the Lee-Parsons approximation $\chi_\Delta(\phi)=(1-3\phi/4)/(1-\phi)^2$, as suggested 
in Ref.~\cite{Meyer2015}, we see from Eq.~\eqref{pt6} and from the observation that $\chi_\Delta(\phi)\geq 1$ that the percolation threshold is systematically 
reduced when compared to Eq.~\eqref{pt5} as many-body effects are taken into account.

\section{Discussion and conclusions}
\label{concl}
By extending the formalism of geometric random graphs to the case of random dispersions of penetrable rods with isotropic orientations, 
we have obtained percolation thresholds that are fully equivalent to those calculated within the second-virial approximation of the 
connectedness Ornstein-Zernike equation \cite{Otten2009,Otten2011}. This result rests on the observation that 
contributions from closed loops of connected rods can be safely neglected for rods with large aspect ratios, so that graph components 
have a tree-like, dendritic structure that allows for a straightforward analytic solution for locating the conditions that correspond to the 
onset of percolation. Our analysis of the case in which rods have an impenetrable hard-core has evidenced that a similar equivalence 
holds between integral equation theory and random geometric graphs. In particular, the second-virial approximation of the connectedness 
Ornstein-Zernike equation corresponds to taking node-degree distributions that are products of Poisson distributions for each orientation of the rods.

We conclude by pointing out that, in contrast to integral equation approaches to continuum percolation \cite{Chatterjee2000,Otten2009}, 
connectedness in random geometric graphs is not restricted to equilibrated and random homogeneous distributions of particles. The present 
formalism is (at least in principle) also applicable to more general systems, in which kinetic and non-equilibrium effects that may reflect the 
processing history of the material modulate the frequency with which contacts are formed between particle pairs, and thereby the percolation 
threshold \cite{Meier2011}.

\appendix
\section{The $n$-cycle coefficient}
\label{appa}

Here we estimate the probability of finding a closed loop of $n$ intersecting rods by considering the $n$-cycle coefficient, which is defined 
as the conditional probability that two rods are connected given that they are the end nodes of an $n$-chain of connected rods. 
For penetrable rods of identical length $L$ and diameter $\Delta$, the $n$-cycle coefficient is given by \cite{Grimaldi2015}
\begin{equation}
\label{a1}
c_n=\frac{\displaystyle\int\prod_{i=1}^nd\vec{r}_id\vec{u}_i\rho(\vec{u}_i)f_{12}f_{23}\cdots f_{n-1,n}f_{n,1}}
{\displaystyle\int\prod_{i=1}^nd\vec{r}_id\vec{u}_i\rho(\vec{u}_i)f_{12}f_{23}\cdots f_{n-1,n}},
\end{equation}
where $f_{ij}=f(\vec{r}_{ij},\vec{u}_i,\vec{u}_j)$ is the connectedness function introduced in Sec.~\ref{mapping} for a pair of penetrable rods with 
orientations $\vec{u}_i$ and $\vec{u}_j$ with their centers separated by $\vec{r}_{ij}=\vec{r}_i-\vec{r}_j$. In writing Eq.~\eqref{a1} we use
the orientational distribution function $\rho(\vec{u})$, introduced in Sec.~\ref{perc1}, which satisfies the normalization condition
$\int d\vec{u}\rho(\vec{u})=1$. Introducing the Fourier transform
$\hat{f}(\vec{q},\vec{u}_i,\vec{u}_j)=\int d\vec{r}f(\vec{r},\vec{u}_i,\vec{u}_j)e^{-i\vec{q}\cdot\vec{r}}$,   Eq.~\eqref{a1} reduces to
\begin{equation}
\label{a2}
c_n=\frac{\displaystyle\int \!\frac{d\vec{q}}{(2\pi)^3}\int\prod_{i=1}^nd\vec{u}_i\rho(\vec{u}_i)\hat{f}(\vec{q},\vec{u}_1,\vec{u}_2)
\cdots\hat{f}(\vec{q},\vec{u}_n,\vec{u}_1)}{\displaystyle\int\prod_{i=1}^nd\vec{u}_i\rho(\vec{u}_i)\hat{f}(0,\vec{u}_1,\vec{u}_2)
\cdots\hat{f}(0,\vec{u}_{n-1},\vec{u}_n)}.
\end{equation}
In the limit of slender rods ($L\gg\Delta$), we calculate the leading-order contribution to $\hat{f}(\vec{q},\vec{u}_i,\vec{u}_j)$  by using the 
coordinate system introduced in Sec.~\ref{mapping}:
\begin{align}
\label{a3}
\hat{f}(\vec{q},\vec{u}_i,\vec{u}_j)=&2L^2\Delta\vert\sin\gamma_{ij}\vert j_0(\vec{q}\cdot\vec{u}_i L/2)j_0(\vec{q}\cdot\vec{u}_j L/2)\nonumber \\
&\times j_0[\Delta\vec{q}\cdot(\vec{u}_i\times\vec{u}_j)/\vert\vec{u}_i\times\vec{u}_j\vert],
\end{align}
where $j_0(x)=\sin(x)/x$. We next introduce the dimensionless momentum defined by $\vec{y}=\vec{q}L/2$ and approximate the last factor 
in Eq.~\eqref{a3} as $j_0[(2\Delta/L)\vec{y}\cdot(\vec{u}_i\times\vec{u}_j)/\vert\vec{u}_i\times\vec{u}_j\vert]\approx 1$ for 
$(2\Delta/L)y\vert\alpha_{ij}\vert\leq 1$ and zero otherwise, where $\alpha_{ij}=\hat{y}\cdot(\vec{u}_i\times\vec{u}_j)/\vert\vec{u}_i\times\vec{u}_j\vert$.
In this way, Eq.~\eqref{a2} can be rewritten as
\begin{equation}
\label{a4}
c_n\approx\frac{\Delta}{L}\tilde{c}_n,
\end{equation}
where
\begin{widetext}
\begin{equation}
\label{a5}
\tilde{c}_n=\frac{2}{\pi^3}\frac{\displaystyle\int\! d\vec{y}\int\prod_{i=1}^nd\vec{u}_i\rho(\vec{u}_i)
\theta\!\left(1-\frac{2\Delta}{L}\alpha_\textrm{max}y\right) j_0(\vec{y}\cdot\vec{u_1})^2\cdots
j_0(\vec{y}\cdot\vec{u_n})^2\vert\sin\gamma_{12}\vert\cdots\vert\sin\gamma_{n1}\vert}
{\displaystyle\int\prod_{i=1}^nd\vec{u}_i\rho(\vec{u}_i)\vert\sin\gamma_{12}\vert\cdots\vert\sin\gamma_{n-1,n}\vert},
\end{equation}
\end{widetext}
where $\alpha_\textrm{max}$ is the maximum value that can be assumed by any of the terms $\vert\alpha_{12}\vert$, $\vert\alpha_{23}\vert$,
$\ldots$, $\vert\alpha_{n1}\vert$. 
For the case of an isotropic orientational distribution of the rods, the denominator of Eq.~\eqref{a5} reduces to $(\pi/4)^{n-1}$. 
Noting that $\vert\sin\gamma_{ij}\vert\leq 1$, we find the following upper bound for $\tilde{c}_n$:
\begin{align}
\label{a6}
\tilde{c}_n &\leq \frac{2^{2n-1}}{\pi^{n+2}}\int\! d\vec{y} \,\theta\!\left(1-\frac{2\Delta}{L}\alpha_\textrm{max}y\right)
\left[\int \frac{d\vec{u}}{4\pi} j_0(\vec{y}\cdot\vec{u})^2\right]^n\nonumber \\
&\approx \frac{2^{2n+1}}{\pi^{n+1}}\int_0^{L/2\Delta}\! dy\, y^{2-n}\left[\int_0^y\! dt\, j_0(t)^2\right]^n,
\end{align}
where we have set $\alpha_\textrm{max}\approx 1$ in the last integral. In the limit of large aspect ratios, the above integral 
becomes proportional to $\ln(L/2\Delta)$ for $n=3$, while it is independent of $L/\Delta$ for $n>3$. We thus obtain that the leading order 
contribution to Eq.~\eqref{a4} scales as
\begin{equation}
\label{a7}
c_n\propto\left\{
\begin{array}{lcc}
\displaystyle\frac{\Delta}{L}\ln\left(\frac{L}{2\Delta}\right)&\textrm{for}&n=3,\\
\displaystyle\frac{\Delta}{L}&\textrm{for}&n>3,\end{array}\right.
\end{equation}
from which we recover the estimate for $n=3$ given in Ref.~\cite{Onsager1949}. Equation \eqref{a7} shows that, for rods with sufficiently large 
aspect ratios, the impact of accounting for such closed loops in calculating the percolation threshold is likely to be minor, and to diminish 
with increasing aspect ratios.

\end{document}